\documentclass[aps,prl,letterpaper,10pt,twocolumn]{revtex4-1}
\usepackage[colorlinks=true,allcolors=blue!60!black]{hyperref}
\usepackage{amssymb}
\usepackage{amsmath}
\usepackage{graphicx}
\usepackage{subfigure}
\usepackage{amsfonts}
\usepackage{xcolor}
\usepackage{epsfig}
\usepackage{color}
\usepackage{bbm}
\usepackage{bm}
\usepackage{tabularx}
\usepackage{multirow}
\usepackage{mathtools}
\usepackage{xfrac}
\newcommand{\ket}[1]{\vert{#1}\rangle}

\newcommand{\abs}[1]{\left|#1\right|}
\newcommand{\xzpm}{z_{\circ,j}}
\newcommand{\gele}{\textsl{g}_{\rm e}}
\newcommand{\Nbos}{\bar{N}}

\newcommand{\tr}[1]{\mathrm{Tr}#1}

\newcommand{\hatd}[1]{\hat{#1}^{\dagger}}

\newcommand{\vnnb}{V$_{\rm N}$N$_{\rm B}$ }

\newcommand{\abinit}{\textit{ab initio} }
\begin{document}
\title{Spin-Mechanical Scheme with Color Centers in Hexagonal Boron Nitride Membranes}

\author{Mehdi Abdi, Myung-Joong Hwang, Mortaza Aghtar, and Martin B. Plenio}
\affiliation{Institut f{\"u}r Theoretische Physik und IQST, Albert-Einstein-Allee 11, Universit{\"a}t Ulm, 89069 Ulm, Germany}

\begin{abstract}
Recently observed quantum emitters in hexagonal boron nitride (hBN) membranes have a potential for achieving high accessibility and controllability thanks to the lower spatial dimension. Moreover, these objects naturally have a high sensitivity to vibrations of the hosting membrane due to its low mass density and high elasticity modulus.
Here, we propose and analyze a spin-mechanical system based on color centers in a suspended hBN mechanical resonator.
Through group theoretical analyses and \abinit calculation of the electronic and spin properties of such a system, we identify a spin doublet ground state and demonstrate that a spin-motion interaction can be engineered which enables ground state cooling of the mechanical resonator. We also present a toolbox for initialization, rotation, and readout of the defect spin qubit. As a result the proposed setup presents
the possibility for studying a wide range of physics. To illustrate its assets, we show that a fast and noise resilient preparation of a multicomponent cat state and a squeezed state of the mechanical resonator is possible; the latter is achieved by realizing the extremely detuned, ultrastrong coupling regime of the Rabi model, where a phonon superradiant phase transition is expected to occur.
\end{abstract}

\maketitle

\begin{figure}[tb]
\includegraphics[width=\columnwidth]{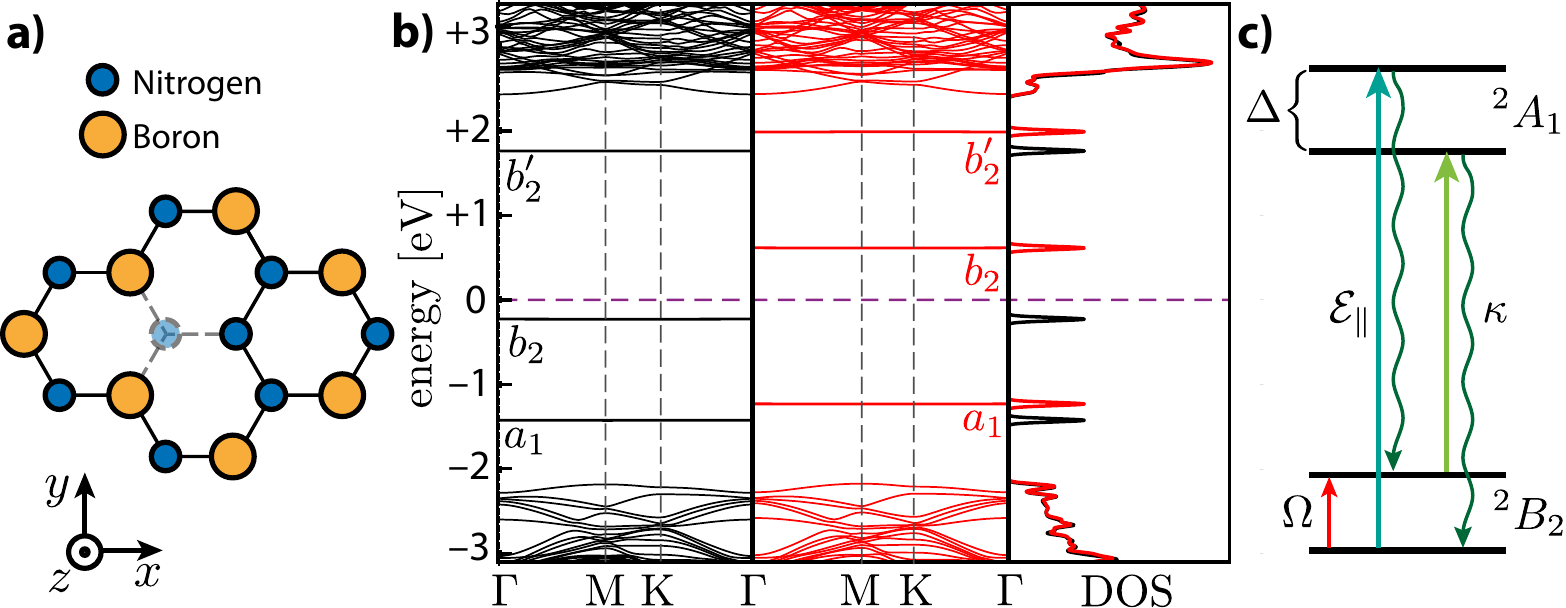}
\caption{%
(a) Geometry of \vnnb defect in hBN monolayer in top view.
(b) \abinit results for band structure and the corresponding density of states (DOS) of a monolayer hBN with charge neutral \vnnb defects: Spin-up polarization is in black and spin-down in red. The defect energy levels within the band-gap are labeled according to their molecular symmetry \cite{suppinfo}. The Fermi energy level is set to zero (purple dashed line).
(c) Two lowest multi-electron states of the defect. They are spin doublets with Zeeman splitting $\Delta$ and the respective transition rates: spin-preserving decays (at rate $\kappa$), spin-flip optical drives due to spin-orbit mixing $\mathcal{E}_\parallel$, and microwave drive $\Omega$. }
\label{fig:model}
\end{figure}

%
%
%----------INTRODUCTION----------%
\textit{Introduction.---}%
Recently, there have been an increasing number of reports on the observation of optical single photon emitters in mono- and multi-layer hexagonal boron nitride (hBN) samples in room and cryogenic temperature setups~\cite{Tran2016a,Jungwirth2016,Chejanovsky2016}. The single-photons emitted from hBN have very narrow and pronounced zero-phonon-line~\cite{Jungwirth2016}. Additionally, the emitters have the privilege of being hosted in a 2D material, giving them the advantage of a high accessibility and sensitivity~\cite{Aharonovich2016,Lovchinsky2017}. The hexagonal boron nitride membranes have low mass yet high elasticity modulus and tensile strength, which make them promising candidates for high quality mechanical resonators~\cite{Lee2010,Song2010,Boldrin2011}. Their extremely small out-of-plane stiffness also gives them a large zero-point amplitude, i.e. a high motion sensitivity.
The origin of hBN emitters has yet to be explored and despite the \abinit computational work in Ref.~\cite{Tran2016a}, which attributes the emissions to charge neutral \textit{anti-site complex} defects \vnnb [Fig.~\ref{fig:model}(a)], a group theory support is still missing.

In this Letter, we perform group theory analyses and \abinit computations to study the electronic structure and spin properties of \vnnb defects.
Based on this knowledge we propose, for the first time, a setup for coupling a spin state of the color centers hosted in a suspended hBN flake to its vibrational modes~\cite{Jin2009,Song2010}. Taking advantage of the exceptional geometry of the proposed setup, a manipulation toolbox for the spin qubit---and through it the mechanical mode---is proposed here. This includes initialization, rotations, and readout of the spin qubit via optical excitations and microwave drives. By employing collimated light beams and tunable microwave drives each spin qubit can be individually addressed.
Regarding the mechanical mode, we shall first demonstrate the possibility of cooling a vibrational mode of the membrane (with frequency $\omega_m$) down to its ground state by the resolved sideband technique.
%($\Gamma \ll \omega_m$) and employing a weakly coupled center $g_0 \lesssim \Gamma$.
We also show that the spin-motion coupling rate $g_0$ can reach and even exceed the fundamental mechanical mode frequency $g_0 \gtrsim \omega_m$, realizing the so-called ultrastrong coupling regime~\cite{Ciuti2006,FornDiaz2010,Niemczyk2010,Yoshihara2016}. Based on such a large coupling rate, we explore the possibility of the fast preparation of nonclassical states of a mechanical mode of the membrane and demonstrate that both multicomponent Schr{\"o}dinger cat states~\cite{Armour2002,Abdi2016} and highly squeezed states can be prepared within less than a mechanical period. While the former makes use of the spin-dependent displacement of the motion, the latter is achieved by effectively realizing the Rabi model in the extremely detuned, ultrastrong coupling regime.
The possibility to realize this regime of the Rabi model opens a door to study the recently predicted superradiant phase transition \cite{Hwang2015,Hwang2016} in this setup providing an alternative to earlier ion trap setups, which rely on interaction picture~\cite{Puebla2017}. The ability to prepare the hBN flake in these nonclassical states can also open an avenue for exploring the macroscopic quantum physics.

%
%
%----------MODEL----------%
\textit{Model.---}%
We consider the \vnnb defect of a monolayer hBN membrane, depicted in Fig.~\ref{fig:model}(a). As group theory analyses predict, consistent with \abinit computations, the ground and excited electronic states---which constitute our manifold of interest---are spin doublets; the results are summarized in Fig.~\ref{fig:model}(b) and c (See the supplemental material for more details and a rigorous treatment \cite{suppinfo}).
Due to dipole allowed transitions, the excited states spontaneously decay to the ground state at the rate $\kappa$ while preserving the spin projection. Meanwhile, a spin-orbit interaction mixes spin states of the two manifolds, inducing spin-flip transitions [the straight lines in Fig.~\ref{fig:model}(c)]. Spontaneously, such transitions are dominated by the spin preserving ones. However, they can be driven by a collimated optical beam with a proper in-plane polarization.
When the setup is immersed into a magnetic field, the spin degeneracies are lifted. We exploit these two properties and propose an optical mechanism for initializing the spin state of the ground manifold \cite{suppinfo}. The induced spin relaxation rate $\Gamma$ is only bounded from above by $\Gamma\leq\kappa$. Hence, it can in principle be switched off---when there is no optical drive, $\mathcal{E}_\parallel=0$. This spin-dependent resonant excitation mechanism also allows single-shot projective readout of the spin \cite{Rogers2014,Robledo2011}.

\begin{figure}[tb]
\includegraphics[width=\columnwidth]{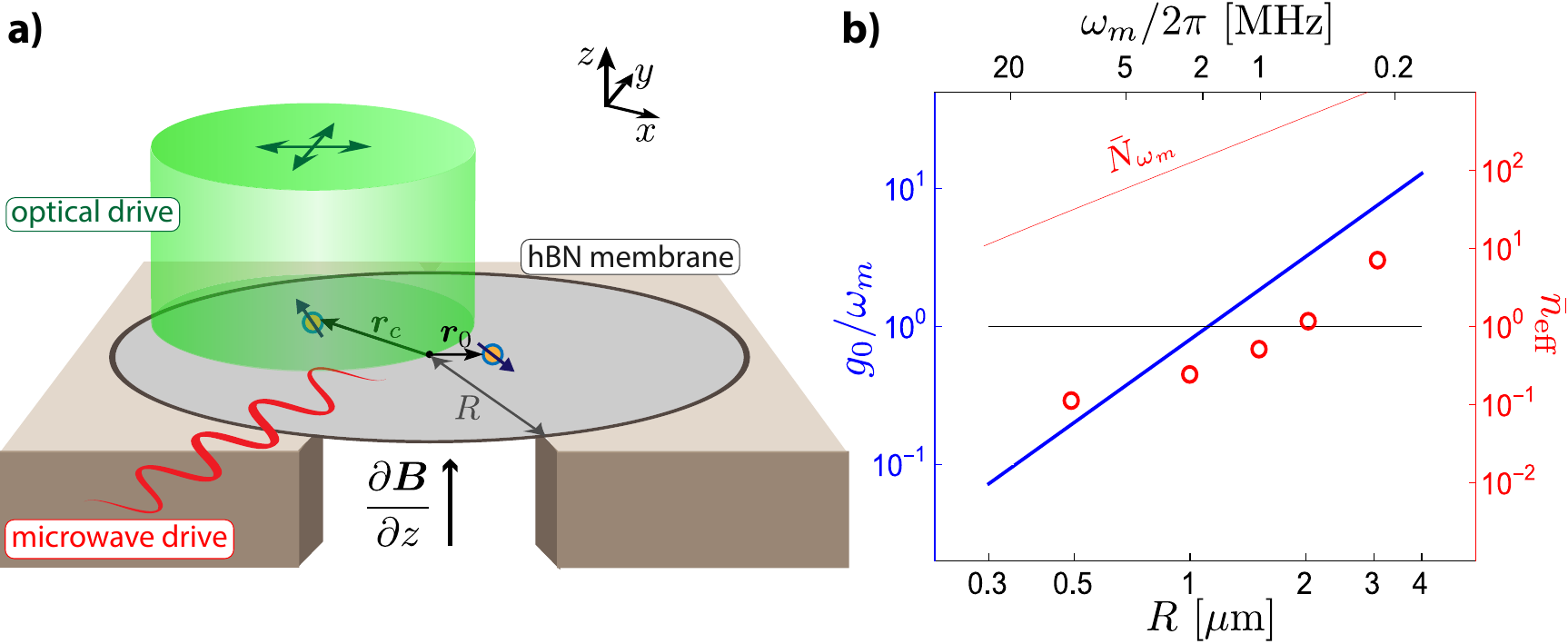}
\caption{%
(a) Scheme for the proposed setup: A suspended circular hBN membrane with radius $R$ as the mechanical resonator. A magnetic field gradient enables a spin-motion coupling. The spin qubits are driven by a microwave field, while a polarized collimated optical drive provides the spin polarization mechanism.
%b) Spin-motion coupling rate as a function of distance from the center of membrane $r$ for different membrane radii $R$. The small red square on each curve denotes optimum $g_c$ (see the text for details).
(b) Coupling rate to mechanical frequency ratio for the qubit at `sweet spot' (blue line) and effective boson number of the mechanical mode after cooling under optimal conditions (red circles) versus radius of the membrane. The black horizontal line indicates both $g_0/\omega_m=1$ and $\bar n_\textrm{eff}=1$ (the ground state limit).
For clarity, the explicit mechanical frequencies are given in the top axis.%
}%
\label{fig:scheme}
\end{figure}

The hBN single-photon emitters have shown intrinsic electronic sensitivity to the local strain~\cite{Jungwirth2016}. However, the observed deformation potential~\cite{Grosso2016} results in very weak zero-point-motion coupling rates (compared to the excited state lifetime) and thus electronic and motional dynamics cannot appreciably influence each other.
Instead, since the ground state of \vnnb color centers is a spin doublet, their electronic spin degree of freedom can couple to the motion of the membrane via a magnetic field gradient~\cite{Wineland1998,Mamin2007,Rabl2009,Grinolds2014,Tao2016} [see Fig.~\ref{fig:scheme}(a) for the proposed setup].
The spin-motion coupling rate depends on the geometry of the membrane, location of the defect, and the geometry of magnetic field.
Hence, it gives the opportunity of investigating various physical phenomena in different working regimes, simply by addressing individual color centers located at different sites.

The prospective high quality factor hBN flake mechanical resonators allow us to single out a mechanical mode. (High quality nano-resonators have already been realized in electro-mechanical experiments for graphene membranes~\cite{Singh2014}). Moreover, assuming weak optical drives ($\mathcal{E}_\parallel \ll \kappa$), the dynamics of the excited states can be adiabatically eliminated and the \vnnb system is described by an effective two-level system, the electronic ground state spin qubit~\cite{Marzoli1994}.
One thus deals with qubits coupled to an isolated mechanical harmonic mode, yet driven and manipulated by classical electromagnetic waves. The following Hamiltonian describes coherent evolution of our system with $N$ spin qubits
\begin{equation}
\hat H_N = \omega_m\hat b^\dag\hat b +\sum_{j=0}^{N-1}\frac{\Omega_j}{2}\hat\sigma_{x,j}-\frac{\delta_{j}}{2}\hat\sigma_{z,j} +g_j\hat\sigma_{z,j}(\hat b+\hat b^\dag),
\label{hamilton}
\end{equation}
where, $\delta_j =\omega_D -\Delta_j$ is the detuning of the microwave drive frequency from the qubit splitting and $\Omega_j$ is the Rabi frequency. Here, $\hat b$ ($\hat b^\dag$) refer to the phonon annihilation (creation) operators of the mechanical mode and $\hat\sigma_x$ and $\hat\sigma_z$ are the Pauli matrices.
The non-uniform magnetic field results in a position-dependent splitting and a spin-motion coupling for each spin qubit of defect.
The splitting $\Delta_j \equiv\Delta(r_j,\theta_j) = \mu_{\rm B} \gele B_\perp(r_j,\theta_j,z=0)$, with $\mu_{\rm B}$ the Bohr magneton and $\gele$ the electron g-factor, is determined by the local magnetic field. The non-uniformity of the magnetic field on the other hand provides the spin-motion coupling, which, in the leading order, is linearly proportional to the mechanical position.
The coupling rates are given by $g_j \equiv g(r_j,\theta_j) = \mu_{\rm B}\gele\partial_z B_\perp(r_j,\theta_j,0)\xzpm$, where $\xzpm = \psi(r_j,\theta_j)\sqrt{\hbar/2m\omega_m}$ is the amplitude of the membrane zero-point-fluctuations at location of the defect with mode profile function $\psi$ and the effective mass $m$ of the mechanical mode. In Hamiltonian (\ref{hamilton}) we have neglected the spin dipole 
interactions between qubits. This is indeed the case when the defects are separated by a 
few hundreds of unit cells.

Fig.~\ref{fig:scheme}(b) presents the coupling rates achievable in our system. Here, the fundamental vibrational mode of a circularly clamped monolayer hBN membrane with radius $R$ with a dominant built-in tensile strain is taken as the mechanical mode \cite{Abdi2016,suppinfo}. Here we consider a maximum magnetic field gradient of $\abs{\partial_zB_\perp(r_0,\theta_0,0)}\approx 270~$G/nm at `sweet spot' $\bm{r}_0=(r_0,\theta_0)$, which can be provided either by a write-head \cite{Tsang2006,Tao2016} or a slightly enhanced magnetic tip~\cite{Mamin2007}.
Remarkably, the maximum spin-motion coupling rate $g_0 \equiv g(r_0,\theta_0)$ can become comparable or even larger than the oscillator frequency $\omega_m$, hence the so-called ultrastrong coupling regime can be achieved [Fig.~\ref{fig:scheme}(b)]. This opens an exciting opportunity to explore the physics of ultrastrong spin-motion coupling with the ground-state cooled mechanical oscillator, as we demonstrate below. Throughout the paper, we will consider only two qubits: one at the field gradient sweet spot, which is used for the ultrastrong spin-motion coupling and one offset by an optimal distance from it for the cooling, as depicted in Fig.~\ref{fig:scheme}(a). The situation can be generalized easily to larger numbers of qubits.
In particular, one can exploit the mechanical mode as a bus for coupling distant spin states to each other, realizing complex spin systems for quantum simulation purposes.

%
%
%----------COOLING----------%
\textit{Cooling.---}%
Now we study the possibility of cooling a mechanical mode of a hBN membrane by a \vnnb color center. The ground state cooling of a mechanical mode coupled to a two-level-system can be achieved for a weak coupling strength and in resolved sideband regime $\Gamma < \omega_m$~\cite{Jaehne2008,Rabl2010,Tian2011}. In our configuration the latter can be satisfied thanks to the tunability of the depolarization rate $\Gamma$. Nevertheless, one also requires a qubit weakly coupled to the resonator in order to cool the resonator to its ground state.
This requirement can be fulfilled by employing an emitter offset from the sweet spot of the membrane where the magnetic field gradient is weaker. The offset $\bm{r}_c-\bm{r}_0$ should be such that the coupling rate $g_c \equiv g(r_c,\theta_c)$ satisfies $g_c<\Gamma$ yet $g_c^2 \gg \Gamma\gamma_m\Nbos_{\omega_m}$. We call such a color center the `cooling qubit'. Then a microwave drive with detuning $\delta_c \approx -\omega_m$ and optimized Rabi frequency $\Omega$ can cool the resonator down to its ground state [Fig.~\ref{fig:scheme}(b)].

The full dynamics of the system is governed by the quantum optical master equation $\dot\rho = \mathcal{L}[\rho]$. The Liouvillian $\mathcal{L}$ includes both coherent and irreversible dynamics of the system. In the weak coupling limit ($g_c \ll \omega_m,\Omega$) the Liouvillian is given by
\begin{align}
&\mathcal{L}[\rho] = -i[\hat{H}_1,\rho] +\frac{\gamma_m}{2}\big((\bar{N}_{\omega_m}+1)\mathcal{D}_{\hat b}[\rho] + \bar{N}_{\omega_m}\mathcal{D}_{\hat b^\dag}[\rho]\big) \nonumber\\
&~~+\frac{\Gamma}{2}\big((\bar{N}_{\Delta_c}+1)\mathcal{D}_{\hat\sigma_{-}}[\rho] + \bar{N}_{\Delta_c}\mathcal{D}_{\hat\sigma_{+}}[\rho]\big)
+\frac{\widetilde\Gamma}{2}\mathcal{D}_{\hat\sigma_{z}}[\rho],
\label{liouv}
\end{align}
where $\gamma_m = \omega_m/Q$ is the mechanical damping rate ($Q$, the mechanical quality factor). $\Gamma$ and $\widetilde\Gamma$ are the qubit relaxation and decoherence rate, respectively. The bosonic thermal occupation number at temperature $T$ is $\bar{N}_\omega = \big(\exp\{\hbar\omega/k_{\rm B}T\}-1\big)^{-1}$ and the Lindblad damping superoperators are $\mathcal{D}_{\hat{o}}[\rho] = 2\hat{o}\rho\hat{o}^\dag -\hat{o}^\dag\hat{o}\rho -\rho\hat{o}^\dag\hat{o}$.
Three main sources of qubit decoherence are considered in our analyses $\widetilde\Gamma = \widetilde\Gamma_o +\widetilde\Gamma_v +\widetilde\Gamma_h$ (see the last section for a discussion); dephasing induced by the optical polarization $\widetilde\Gamma_o$ (which is comparable to the relaxation rate $\widetilde\Gamma_o\approx\Gamma$), the one stemming from membrane vibrations $\widetilde\Gamma_v$, and pure dephasing as a result of the hyperfine interaction with nuclear spin bath $\widetilde\Gamma_h$~\cite{Hanson2007}.
We numerically solve the steady state of the master equation $\rho_{\rm ss}$ with parameters: $Q=1\times 10^5$, $T=14$~mK, $\Gamma=\widetilde\Gamma_o=\widetilde\Gamma_v=0.1\omega_m$, $\widetilde\Gamma_h \approx 2\pi \times 100$~kHz, and other parameters are optimized for the best cooling. The mean mechanical phonon numbers $\bar{n}_{\rm eff}=\tr\{\hat b^\dag\hat b\rho_{\rm ss}\}$ for several $R$ values are plotted in Fig.~\ref{fig:scheme}(b). Remarkably, the mechanical mode can be cooled down to the ground state even for cases with large $\Nbos_{\omega_m}\gg 1$ [see the thin red line in Fig.~\ref{fig:scheme}(b)].

%

\begin{figure}[tb]
\includegraphics[width=\columnwidth]{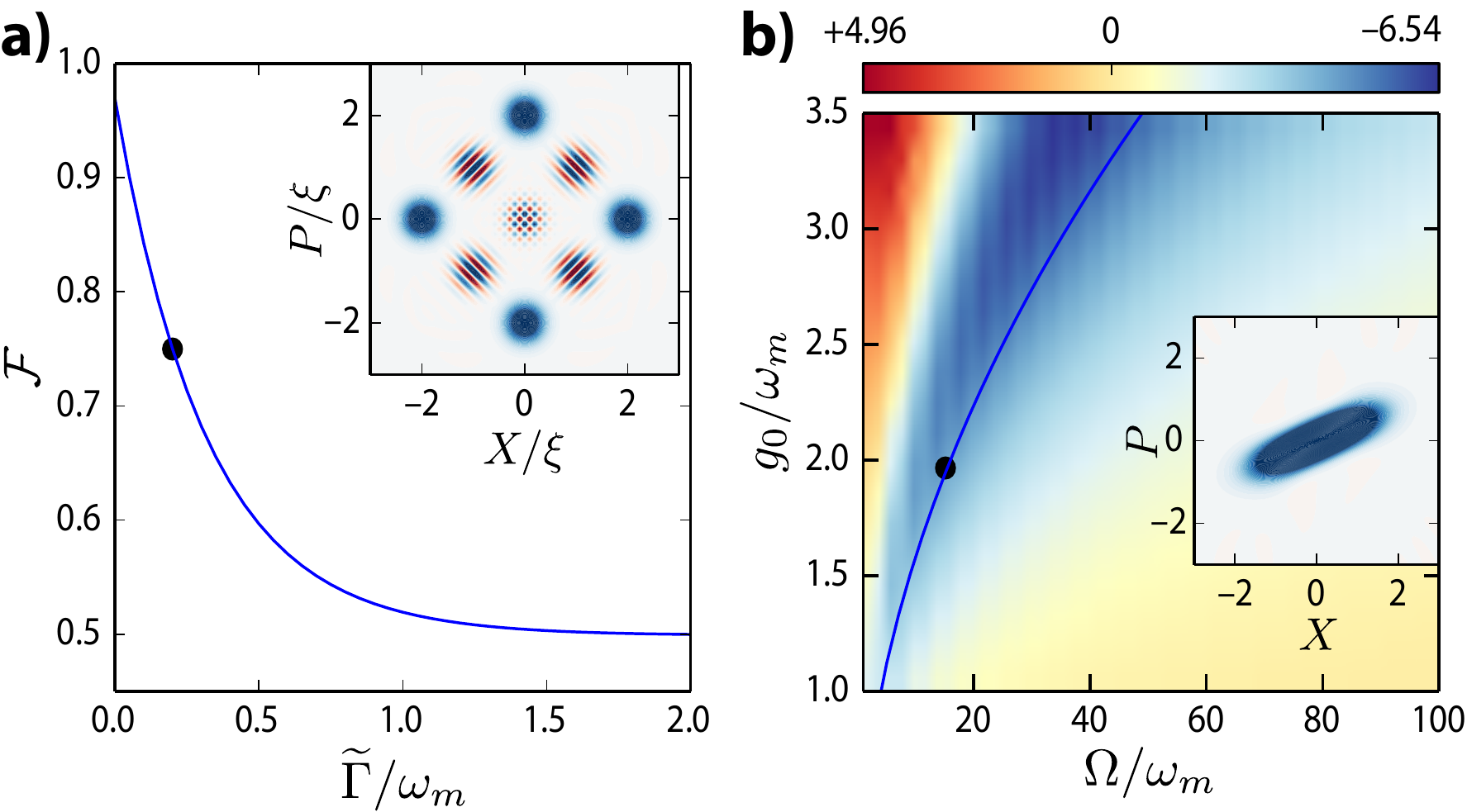}
\caption{%
(a) The fidelity of the prepared multicomponent Schr\"odinger cat state as a function of decoherence rate. Inset: The Wigner function of the prepared state for $\widetilde\Gamma/\omega_m=0.2$ (black dot).
(b) The minimum quadrature variance of the mechanical mode in units of dB. Inset: The Wigner function of the prepared state for $\xi=2$ and $\Omega/\omega_m=15$ (black dot). Note that in both plots $\widetilde\Gamma_o \simeq 0$ because there is no optical polarization and $\widetilde\Gamma_h/\omega_m \approx 0.1$.
}
\label{fig:state}
\end{figure}

%
%
%----------STATE----------%
\textit{Multicomponent cat state.---}%
Here, we first note that even in the presence of a central qubit, which realizes the ultrastrong coupling with the mechanical mode, the ground state cooling can still be achieved using the cooling qubit. Owing to the $r$ dependence of splitting $\Delta(r)$, the central qubit is off-resonant with respect to the microwave field employed for cooling $\Delta_0 \gg \Omega$; consequently, it does not significantly alter the cooling dynamics of the system~\cite{suppinfo}. Therefore, it is possible to cool down the mechanical mode to its ground state, and then to employ the central qubit for an ultrafast preparation of nonclassical states of the membrane.
As an example, we put forth a protocol to prepare a multicomponent cat state of the mechanical mode, which we pursue due to its beneficial properties for continuous variable processing and metrology~\cite{Vlastakis2013,Zurek2001}. The protocol is following:
i) Once the mechanical mode is cooled down to its ground state, the spin qubit is initialized in $\ket\downarrow$ by the optical polarization, followed by a microwave $\pi/2$-pulse to create a spin superposition.
ii) After a quarter mechanical period, during which the spin-dependent displacement occurs, another $\pi/2$-pulse is applied.
iii) After another quarter mechanical period, a $\pi/2$-pulse is applied, followed by the optical readout of the qubit state. The postselected mechanical state is
$\frac{1}{2} \Big[e^{2i\xi^2}\big(e^{-i\phi}\ket{2\xi}\pm e^{i\phi}\ket{-2\xi}\big) - e^{-2i\xi^2}\big(\ket{-2i\xi} \mp\ket{2i\xi}\big)\Big]$. Here, the upper (lower) sign corresponds to the measurement outcome of spin up (down), $\xi =g_0/\omega_m$ is the spin-dependent displacement, and $\phi = \pi\Delta_0/2\omega_m$ is the accumulated phase during the first half of a mechanical period. The fidelity of the prepared state following the above protocol for a membrane of radius $R=1.5~\mu$m, which leads to a spin-dependent displacement of $\xi \approx 2$, is presented in Fig.~\ref{fig:state}(a) as a function of the qubit decoherence rate.
Thanks to the ultrafast preparation in less than a mechanical period, the prepared state is noise-resilient, as evidenced by a clear interference pattern of the Wigner function in the presence of the sizable decoherence rate [Fig.~\ref{fig:state}(a)]. We also note that one can prepare mechanical Fock states of the membrane by employing a protocol proposed in Ref.~\cite{Abdi2015}.
%

%
%
%----------RABI----------%
\textit{The Rabi model and squeezed state.---}%
The Hamiltonian $\hat H_1$ in Eq.~(\ref{hamilton}) with a central qubit can be transformed into the Rabi Hamiltonian by setting $\delta_0=0$ and applying a spin rotation unitary transformation: $\hat H_{\rm R} = (\Omega/2)\hat\sigma_z -g_0\hat\sigma_x(\hat b +\hat b^\dag) +\omega_m \hat b^\dag\hat b$. We emphasize that $\Omega$ is set by the microwave Rabi frequency, whose magnitude can be controlled independent of $g_0$ and $\omega_m$ and is only limited by the maximum input power that does not heat up the system. Therefore, a broad parameter regime of the  Rabi model encompassing the ultrastrong and deep strong coupling regime~\cite{Yoshihara2016,Niemczyk2010, FornDiaz2010}, as shown in Fig.~\ref{fig:scheme}(b), both in a resonant or dispersive limit can be explored here.

A particularly interesting achievable limit is $\Omega \gg g_0 \gg \omega_m$, in which finite-component systems of coupled spins and bosons, including the Rabi model, have been shown to undergo a superradiant phase transition~\cite{Hwang2015,Hwang2016}.
In the zeroth order in $\omega_m/\Omega$, one adiabatically eliminates the spin excited state to arrive at an effective Hamiltonian, $\hat H_\textrm{eff}=\omega_m \hatd b \hat b - (g_0^2/\Omega) (\hat b+\hatd b)^2 +\mathcal{O}(\omega_m/\Omega)$~\cite{Hwang2015}. For coupling strength $g_0=\sqrt{\omega_m\Omega}/2$ the $\hat H_\textrm{eff}$ becomes unstable marking the critical point of the Rabi model in the limit of $\omega_m/\Omega\rightarrow 0$. For finite, but large $\Omega/\omega_m$, which can be realized in our setup, even though the higher order terms can no longer be neglected the short time dynamics is largely dominated by $\hat H_\textrm{eff}$. Noting that $\hat H_\textrm{eff}$ includes the so-called single-mode squeezing terms, $\hat b^{\dagger 2}$ and $\hat b^2$, we propose to use the strong microwave drive realizing the Rabi model to generate a squeezed state of the membrane: After the ground state cooling and the spin initialization, a resonant ($\delta_0=0$) microwave field with a Rabi frequency $\Omega$ is applied over a quarter of the mechanical period. By a projective readout of the qubit, a mechanical squeezed state will be prepared conditioned on the spin down outcome.

We simulate the protocol in the presence of decoherence and numerically calculate the variance of mechanical quadratures, $\hat X(\theta)\equiv \hat be^{i\theta}+\hatd b e^{-i\theta}$, of the prepared states for a wide range of achievable values of $\xi$ and $\Omega$. The minimum variance values are presented in Fig.~\ref{fig:state}(b) in units of decibel. It shows a significant amount of squeezing, up to around $-6.5$~dB.
The maximum squeezing is achieved for parameters close to the critical point $4g_0^2=\omega_m\Omega$ [the blue line in Fig.~\ref{fig:state}(b)]. The Wigner function of a final state clearly demonstrates that a squeezed state is indeed prepared. Note that we have used the master equation in the dressed basis in our numerical calculation~\cite{Beaudoin2011}.

%Moreover, starting from mechanical vacuum spin-down state, a squeezed vacuum as a result of a quarter Rabi period $\Delta\hat x \approx 1.23$ and $\Delta\hat p \approx 0.787$.

%
%
%----------EXTRA----------%
\textit{Discussion.---}%
Throughout the paper we have carried out simulations under the assumption that at each step only one of the qubits effectively manipulates the membrane and the other one remains idle. This indeed is justifiable because the cooling qubit is weakly coupled to the membrane and the magnetic field splitting $\Delta(r)$, and hence the detuning, for each qubit is different from one another  which enables selective resonant driving~\cite{suppinfo}.
%The only exception to this may happen in the cooling process, where---even though off-resonant---the strongly coupled central qubit may hinder ground state cooling. We have nonetheless shown in the supplemental information and numerically confirmed that its sole effect is imposing a displacement in the mechanical phase space. So, the mean thermal phonon number of the mechanical mode equals those reported before.

For the spin qubit, possible intrinsic sources of decoherence include coupling of the spin to lattice vibrations, hyperfine interaction of the electron spin with the surrounding nuclear spins \cite{deSousa2003}, noise in the external magnetic field \cite{Roos1999}, and spin dipole interaction with the other defects.
The hyperfine induced relaxation rate is estimated to be negligible here as we assume splittings much larger than the hyperfine energies \cite{Hanson2007}. Its pure dephasing contribution is taken into account in our analysis. Here, it worth mentioning that $\widetilde\Gamma_h \simeq 0$ if the nuclear spin lattice is hyperpolarized~\cite{Abragam1978}.
Moreover, the phonon DOS of hBN only exhibits large contributions at high frequencies \cite{Serrano2007}.
At the low temperatures that we are considering here only the lowest vibrational modes can contribute in the decoherence with rate $\widetilde\Gamma_v$.
%The magnetic field fluctuations also lead to negligible dephasing.
As an extrinsic source, the optical polarization process also leads to a qubit dephasing, that is comparable to the relaxation rate in size.
%This affects only the `cooling qubit', which requires a continuous polarization, $\widetilde\Gamma_o \simeq \Gamma$, and the `central qubit', operating without a spin polarization, hence gets a higher coherence time $\widetilde\Gamma_o \simeq 0$.

The proposed spin-mechanical setup offers the potential for a broad range of applications, which include ultrasensitive force detection~\cite{Moser2013,Muschik2014}. Also, the nonzero nuclear spins of nitrogen and boron stable isotopes give the opportunity of using the hBN flakes as a platform for quantum simulation of 2D spin systems, where the electronic spin of the color centers will realize their initialization, control, and readout~\cite{Cai2013}.

%
%
%----------ACKNOLEDGEMENT----------%
\begin{acknowledgements}
\textit{Acknowledgements.---}%
This work was supported by the Alexander von Humboldt foundation, the ERC Synergy grant BioQ, the EU STREP project EQUAM, the DFG CRC TRR21, and DFG FOR 1493.
The authors also acknowledge support by the state of Baden-W{\"u}rttemberg through bwHPC and the German Research Foundation (DFG) through grant no INST 40/467-1 FUGG.	
\end{acknowledgements}

\vspace{-5mm}

%
%
%----------REFERENCES----------%
\bibliography{spinomechanics}

\end{document}